\documentstyle[aps,twocolumn,prl,epsf]{revtex}
\begin{document}
\title{Evidence for exhaustion in the conductivity of the 
infinite-dimensional periodic Anderson  model}
\author{A.N.\ Tahvildar-Zadeh$^1$, M.\ Jarrell$^1$, 
Th.\ Pruschke$^2$ and J.\ K.\ Freericks$^3$ }
\address{$^1$Department of Physics, University of Cincinnati, Cincinnati,
Ohio\\
$^2$ Institut f\"ur Theoretische Physik, Universit\"at Regensburg,
Regensburg, Germany\\
$^3$Department of Physics, Georgetown University, Washington, DC 20057
}
\date{\today}
\maketitle
\begin{abstract}
        Monte Carlo-maximum entropy calculations of the conductivity of 
the infinite-dimensional periodic Anderson model are presented.  We show 
that the optical conductivity displays anomalies associated with the exhaustion 
of conduction-band states near the Fermi energy including a Drude weight 
which rises with temperature, with weight transferred from a temperature 
and doping-dependent mid-IR peak and a low-frequency incoherent contribution.
Both the Drude and mid-IR peaks persist up to very high temperatures. 
The resistivity displays a non-universal peak and two other regions 
associated with impurity-like physics at high temperatures and Fermi Liquid 
formation at low $T$.
\end{abstract}
%\renewcommand{\thefootnote}{\copyright}
%\footnotetext{ 1994 by the authors.  Reproduction of this article by any means
%is permitted for non-commercial purposes.}
%\renewcommand{\thefootnote}{\alpha{footnote}}


\section{Introduction}

Metallic compounds containing rare earth elements with partially filled $f$ 
shells, such as CeBe$_{13}$ or UPt$_{3}$, belong to the general category of 
heavy fermions materials\cite{review}. They are  characterized by a large Pauli 
susceptibility and  specific heat as compared to ordinary metals, which 
indicate a large effective electronic mass, and also by anomalous transport 
properties such as non-monotonic temperature dependence of the resistivity. 
These anomalies are usually attributed to the formation of a resonant 
state at the Fermi energy, associated with moment screening, due to the 
admixture of the electronically active and well localized $f$ orbitals with 
the metallic band of the host.  As the Kondo screening peak develops, the optical 
conductivity develops a pronounced low-frequency Drude peak.  Interband 
transitions across the Fermi energy can also yield strongly 
temperature-dependent features in the mid-IR optical response.

Heavy fermion (HF) materials are usually modeled by the periodic 
Anderson model (PAM) or the single impurity Anderson model (SIAM). 
While the SIAM is able to capture the general physics of moment
formation and subsequent Kondo screening and is thus sufficient to 
understand most of the thermodynamics of HF materials,  lattice effects 
like coherence and correlations between different sites are of course
completely neglected in such a simplified picture.
Coherence effects lead for example to a vanishing low-temperature resistivity
in the metallic regime and to the formation of a band insulator in the
symmetric PAM.
Correlations between the sites on the other hand are responsible
for various types of phase transitions which lead to an extremely
rich phase diagram with metallic, insulating or
near insulating regimes  depending
upon the filling and model parameters.

In the metallic regime of the PAM, perhaps the most important but also
controversial lattice effect is the so called exhaustion as proposed
by Nozi$\grave{e}$res.  The ``exhaustion'' 
problem~\cite{noz1,noz2} occurs when a {\it few} mobile electrons, 
$n_{scr}$, have to screen {\it many} local moments, $n_f$, 
in a metallic environment, i.e, $n_{scr} << n_f$. This situation is 
engendered by the fact that only the electrons within $T_K$ (where $T_K$ 
is the single impurity Kondo temperature) of the Fermi surface can 
effectively participate in screening the local moments.  Thus the  
number of screening electrons can be estimated as $n_{scr}=N_d(0) T_K$, 
where $N_d(0)$ is the conduction band density of states at the Fermi 
level.  A measure of exhaustion is the dimensionless ratio
$p=\frac{n_f}{n_{scr}}=\frac{n_f}{N_d(0) T_K}$ \cite{noz1}.
Nozi$\grave{e}$res has argued that $p$ is roughly the number of scattering 
events between a local moment and
a mobile electron necessary for the mobile electron's spin to precess 
around a local moment by $2\pi$.\cite{noz2} In the 
case $p\gg 1$, when the number of screening electrons is much smaller than 
the number of local moments to screen, magnetic screening is necessarily 
collective and the single impurity picture becomes invalid. 

One further consequence of Nozi$\grave{e}$res' idea then is \cite{noz1} that
the small fraction $n_{scr}$ of screened states may be viewed as relatively
stable polaron-like particles at low 
temperatures $T<T_K$.  He thus proposed that the screened and unscreened sites 
may be mapped onto particles and holes of an effective single-band Hubbard 
model, respectively. The screening clouds can hop from site to 
site and may effectively screen all the moments in a dynamical fashion.

In previous work, we extended \cite{niki_prl} this argument to explain the 
strong reduction of the Kondo scale \cite{niki2} in the metallic regime 
of the PAM. The hopping constant of Nozi$\grave{e}$res' effective model is 
suppressed relative to the bare one by the overlap of the screened and 
unscreened states.  This, together with the fact that the 
filling of the effective Hubbard model is very close to half filling
($n_{eff}\approx n_f-n_{scr}\approx 1$ for $n_f=1$), 
suggests that the relevant low-energy 
scale $T_0$ of the effective model becomes very small, compared to the Kondo 
impurity temperature $T_K$. 

In the crossover regime characterized by temperatures between $T_K$, where 
conventional Kondo screening begins, and the lattice scale $T_0$, where
coherence forms, the quasiparticle peak and the screened local moment evolve 
much more slowly than their counterparts in the SIAM\cite{niki_prl}.  This 
provides a possible explanation for the slow evolution of the quasiparticle 
peak seen in photoemission experiments\cite{jjaa}.  However, these results 
remain controversial\cite{debate} and are complicated by the fact 
that photoemission is very surface sensitive\cite{allen_ybsurface}, and probes 
only the first $\approx 10\AA$ of the surface under investigation.  Thus, 
predictions for less surface sensitive probes such as transport and optical 
conductivity are important.

In this paper we study the effect of screening on the optical conductivity 
and resistivity of the PAM.  We find that the resistivity is characterized 
by an impurity-like regime at high temperatures $T\approx T_K$, a crossover
regime with non-universal behavior, and a low temperature regime $T\alt T_0$ 
where coherence begins to form.  The crossover regime is the most interesting
and is characterized by the exhaustion of screening states near the Fermi 
surface, as measured 
by the development of a dip centered at the Fermi energy, $\omega=0$, in 
the effective f-d hybridization $\Gamma(\omega)$.  The peak in the 
resistivity, which marks the upper end of the crossover regime, is a 
non-universal feature centered at a temperature $T_e$, $T_0<T_e<T_K$.
$T_e$ is the temperature at which exhaustion first becomes apparent
as a reduction in $\Gamma(0)$.  In the exhaustion region, the Drude weight 
increases strongly with temperature, and the lost weight at low $T$ is 
largely transfered to a mid infra-red (MIR) peak.  
This feature is due to interband transitions and its position matches
with the minimum of the direct energy gap between the two quasi-particle
bands below and above the Fermi energy. 
Both the Drude and MIR peaks persist up to unusually high temperatures
$T\approx 10 T_0$.  We interpret the persistence of these features 
as well as the unusual decline of the Drude weight as $T\to 0$ as 
manifestations of Nozi$\grave{e}$ires' exhaustion scenario.

\section{Formalism} 

The Hamiltonian of the PAM on a $D$-dimensional hypercubic lattice is
\begin{eqnarray}
H &=& \frac{-t^*}{2\sqrt{D}}\sum_{\langle ij\rangle \sigma}
\left ( d^\dagger_{i\sigma}d_{j\sigma}+{\rm H.c.}\right )\nonumber \\
&+&
\sum_{i\sigma}\left(
\epsilon_{d}d^\dagger_{i\sigma}d_{i\sigma}+
\epsilon_{f}f^\dagger_{i\sigma}f_{i\sigma}
\right)
+V\sum_{i\sigma}\left(d^\dagger_{i\sigma}
f_{i\sigma}+{\rm H.c.}\right)\nonumber \\
&+&\sum_{i} U(n_{fi\uparrow}-1/2)(n_{fi\downarrow}-1/2)\;\;\label{Ham}.
\end{eqnarray}
In Eq. (1), $d(f)^{(\dagger)}_{i\sigma}$ destroys (creates) a $d(f)$ 
electron with spin $\sigma$ on site $i$. The hopping is restricted to 
the nearest neighbors and scaled as $t=t^*/2\sqrt{D}$.  We take $t^*=1$
as our unit of energy.  $U$ is the 
screened on-site Coulomb repulsion for the localized $f$ states and 
$V$ is the hybridization between $d$  and  $f$ states. This model 
retains the features of the impurity problem, including moment formation 
and screening, but is further complicated by the lattice effects. 

Metzner and Vollhardt \cite{mevoll} observed that the irreducible 
self-energy and vertex-functions become purely local as the coordination 
number of the lattice increases.  As a consequence, the solution of an 
interacting lattice model in $D=\infty$ may be mapped onto the solution of 
a local correlated impurity coupled to a self-consistently determined 
host\cite{infdrev}.  We employ the quantum Monte Carlo (QMC) 
algorithm of Hirsch and Fye \cite{fye} to solve the remaining
impurity problem and calculate the  imaginary time local Green's functions. 
We then use  the maximum entropy method (MEM)\cite{MEMlong} to find the $f$ 
and $d$ density of states  and  the  self-energy\cite{symmpam}.
In the limit of $D=\infty$ the two-particle vertex corrections to the optical 
conductivity ($\sigma(\omega)$) vanish identically due to symmetry restrictions
\cite{Khurana}.
Hence $\sigma(\omega)$ can be calculated by the knowledge of one-particle
self energy through a particle-hole bubble diagram which includes the 
appropriate electron-photon interaction vertices $e{\bf v_k}$ at
each end. A straightforward calculation of this diagram leads
to\cite{Schweitzer}, 
\begin{eqnarray}
\sigma_{xx}(\omega)&=&{e^2\pi\over {\cal V}D} \int_{-\infty}^{+\infty}
d\epsilon 
{f(\omega)-f(\epsilon+\omega)\over \omega} \nonumber \\ 
& &\int_{-\infty}^{+\infty} dy
\rho(y)A_d(y,\epsilon)A_d(y,\epsilon+\omega),
\label{opticalconductivity}
\end{eqnarray}
where ${\cal V}$ is the lattice volume, $f$ is the Fermi function, 
$A_d(\epsilon_k,\omega)=-{1\over\pi}Im[G_d(k,\omega)]$ is the conduction 
band spectral function and $\rho(y)=exp(-y^2)/\sqrt{\pi}$ is the noninteracting
density of states.  

\section{Results}  

We calculated the transport for a variety of model parameters.  The QMC-MEM 
results are limited to relatively small values of $U/V$ by the QMC procedure 
and relatively low temperatures where MEM may be used to calculate the 
dynamics.  Here we present QMC results for $U=1.5$, $V=0.6$, $n_f\approx 1$ 
for three conduction-band fillings $n_d=0.4,\,0.6$, and $0.8$. 
The impurity scale $T_K$ was calculated by extrapolating the 
local susceptibility to $T=0$ (i.e. $T_K=1/\chi_{imp}(T\to 0)$).
The lattice Kondo temperature was similarly calculated by extrapolating the 
{\em effective} local susceptibility of the PAM which is the additional 
local susceptibility due to the introduction of the effective impurity into 
a host of d-electrons\cite{symmpam}.
These  extrapolations result $T_K=0.2, 0.46, 0.5$
and $T_0=0.017, 0.055, 0.149$ for the three band fillings mentioned above,
respectively.

\subsection{Optical Conductivity}

\begin{figure}[htb] 
\epsfxsize=3.0in
\epsffile{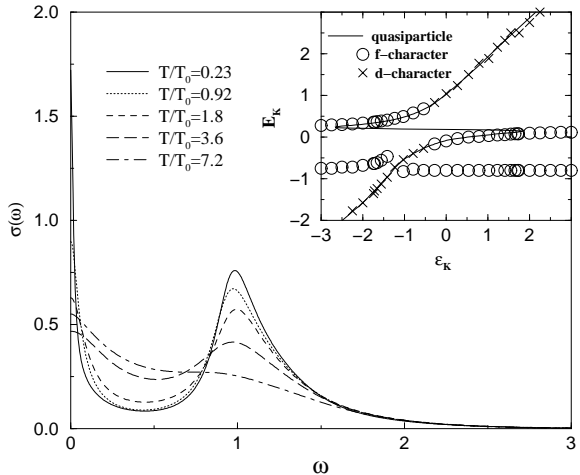}
\caption[]{\em {Optical conductivity  for various temperatures when $U=1.5$, 
$V=0.6$, $n_f\approx 1$, $n_d=0.6$ and $T_0=0.055$.  Both Drude and
Mid-IR peaks are present and persist up $T/T_0 \agt 10$.  The inset shows the 
corresponding band energies $E_k$ plotted vs.\ the bare d-band dispersion 
$\epsilon_k$ when $T/T_0=0.46$.  The solid line shows the quasiparticle energy 
calculated as the real part of the Green's functions poles. The symbols show 
the positions of the maxima in the $f$ and $d$ spectral functions.  The mid-IR 
peak in the optical conductivity is due to direct transitions between occupied 
and unoccupied quasiparticle states. 
}}
\label{Optical_Band}
\end{figure}

The real part of the optical conductivity calculated with QMC-MEM is plotted 
versus frequency in Fig.~\ref{Optical_Band} for several temperatures.  As 
the temperature is lowered $T\alt T_K$ two features begin to develop, a 
Drude peak at zero frequency, associated with free quasiparticles, and 
a Mid-IR peak at $\omega\approx 1$ generally associated with an interband 
transition as we will discuss later.  

\begin{figure}[htb] 
\epsfxsize=3.0in
\epsffile{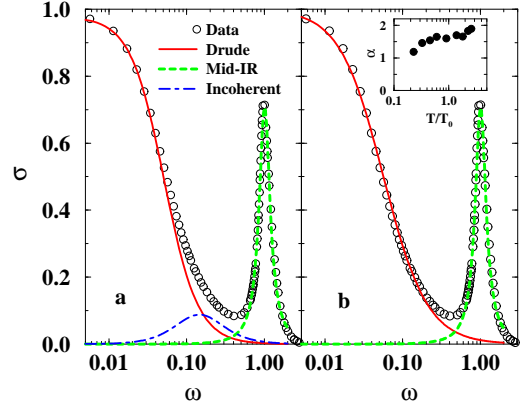}
\caption[]{\em {
Two fitting schemes for separating the share of Drude and interband transition 
to the optical conductivity. The model parameters are the same as in the 
previous figure and $T/T_0=0.46$. a) Fitting to Eq.~\ref{opt_cond_fit}, a 
Fermi-liquid Drude form and two harmonic-oscillator forms. One of the 
harmonic-oscillators accounts for the interband transitions in the mid-IR 
region and the other for the  low frequency spectral weight that remains 
incoherent. b) Fitting to Eq.~\ref{nflfit}, a ``generalized'' Drude form
for the low frequency part and a  harmonic-oscillator for the mid-IR
region. Using the exponent in the ``generalized'' Drude form as a fitting
parameter (shown in the inset) renders the use of an incoherent 
harmonic-oscillator peak unnecessary.
}}
\label{fitschemes}
\end{figure}

To be consistent with the way that experimental data is usually 
analyzed the features in $\sigma(\omega)$ are fit to a Lorentzian plus two
(asymmetric) harmonic-oscillator forms for the higher-energy peak 
\begin{eqnarray}
 {\sigma(\omega)}&\approx& \frac{D_0}{\pi} 
\frac{\tau}{1+\omega^2\tau^2} 
+ \frac{C_{MIR}}{\pi} 
\frac{\omega^2\Gamma_{MIR}}{\omega^2\Gamma^2_{MIR}+(\omega^2-\omega_{MIR}^2)^2}
\nonumber \\
&+& \frac{C_{inc}}{\pi} 
\frac{\omega^2\Gamma_{inc}}{\omega^2\Gamma^2_{inc}+(\omega^2-\omega_{inc}^2)^2}
,
\label{opt_cond_fit}
\end{eqnarray}
with $\tau$ the relaxation time of the ``free'' quasiparticles, and the 
constants $C_{MIR}$, $\omega_{MIR}$, and $\Gamma_{MIR}$ are respectively 
the weight, center, and width  of the mid-IR peak.  As shown in 
Fig.~\ref{fitschemes}(a), in order to improve the fitting procedure we   
include a second harmonic-oscillator form centered at 
$\omega_{inc} \approx 0.1$.  Since this weight cannot be fit to the Drude 
form, part of the low frequency spectral weight remains incoherent as $T$ 
is lowered.  An alternative method to separate the contributions of the 
Drude and the mid-IR peak to the  spectral weight is to fit the optical 
conductivity data to a ``generalized'' Drude form and  a 
harmonic-oscillator with strength $C^*_{MIR}$ for the mid-IR region as 
discussed above
\begin{equation}
{\sigma(\omega)} \approx
\frac{D^*}{\pi} \frac{\tau}{1+(\omega\tau)^{\alpha}} +
\frac{C_{MIR}^*}{\pi} 
\frac{\omega^2\Gamma_{MIR}^*}{\omega^2\Gamma^{*2}_{MIR}+
(\omega^2-\omega_{MIR}^{*2})^2}\,.
\label{nflfit}
\end{equation}
For a generic Fermi-liquid model $\alpha =2$ as was 
assumed in the previous fitting form. But, it turns out that treating $\alpha$ 
as a fitting parameter renders the use of incoherent harmonic-oscillator 
unnecessary as shown in Fig.~\ref{fitschemes}. The inset to this figure
shows the value of exponent $\alpha$ as a function of $T$.

\begin{figure}[htb] 
\epsfxsize=3.0in
\epsffile{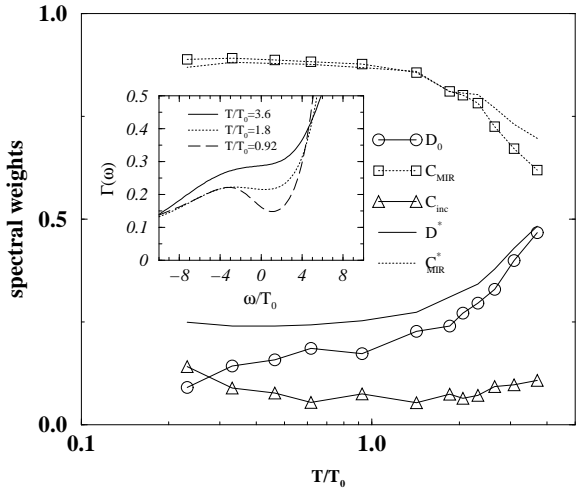}
\caption[]{\em {The Drude, mid-IR and incoherent weights  vs.\ 
temperature  when $U=1.5$, $V=0.6$, $n_f\approx 1$, $n_d\approx 0.6$ and
$T_0=0.055$.  The coefficients $D_0$, $C_{MIR}$ and $C_{inc}$ (
$D^*$ and $C^*_{MIR}$) were
determined from a fit to Eq.~\ref{opt_cond_fit} (Eq.~\ref{nflfit}).
The inset shows the effective f-d hybridization 
$\Gamma(\omega)={\rm{Im}}\left(1/G_f(\omega)+\Sigma(\omega)\right)$ plotted
vs.\ $\omega$ for several different temperatures.
The weight in the Drude peak falls with $\Gamma(0)$ since the weight
in the quasiparticle peak is expected to vary as
$\sim \exp(-\pi U/8\Gamma(0))/\Gamma(0)$ (see text).  
}}
\label{Drude_Gamma}
\end{figure}

The spectral weights $D_0$, $C_{MIR}$, $C_{inc}$, as well as those from
the ``generalized'' fit $D^*$ and $C^*_{MIR}$ are plotted in  
Fig.~\ref{Drude_Gamma} when $U=1.5$, $V=0.6$, $n_f\approx 1$ and  $n_d=0.6$ 
($T_0=0.055$). The most surprising result is that there is a significant
transfer of weight from high to low frequencies with increasing temperature.
The weight in the Drude peak, as measured by either $D_0$ or $D^*$, rises 
with temperature at low $T$.  This feature was independent of the fitting 
procedure applied and was even found if the extrapolation technique of 
Scalapino {\em{et al.}}\cite{swz} was employed at low temperatures $T<T_K$.  
The opposite trend is generally expected since the Drude peak is associated 
with the buildup of quasiparticle weight in the single-particle spectra at 
low temperatures.  

To interpret the drop in the Drude weight with lowering $T$, we calculate 
the effective hybridization strength $\Gamma(\omega)$,
\begin{equation}
\Gamma(\omega) =
{\rm{Im}}\left(\Sigma(\omega)+{1\over G_f(\omega)}\right),
\end{equation}
where $G_f(\omega)$ is the local $f$ Green's function and
$\Sigma(\omega)$ is the local f-electron self-energy.
$\Gamma(\omega)$ is a measure of the hybridization between the effective
impurity in the DMF problem and its medium (for example, in the SIAM
$\Gamma(\omega)= \pi V^2 N_d(\omega)$, where $N_d(\omega)$
is the d-band density of states).  The inset to Fig.~\ref{Drude_Gamma} shows 
the effective hybridization for the PAM near the Fermi energy.
$\Gamma(\omega)$ begins to develop a dip at the Fermi energy at roughly the
same temperature where Drude weight begins to drop.   

The formation of this dip in $\Gamma(\omega)$ can be interpreted as an
effective reduction or "exhaustion" of the states near the Fermi energy 
responsible for screening the local moments. In infinite dimensions, other
mechanisms which could be responsible for the dip, such as non-local spin 
or charge correlations, are absent\cite{Raja}.  We therefore believe that 
it can be taken as direct evidence for Nozi$\grave{e}$res' exhaustion 
scenario. Concomitant with the development of this dip is a drop in the 
quasiparticle weight as may be seen from the following simple qualitative 
argument:
%The density of states of the
%lattice is equivalent to the DOS of the effective impurity problem.
The relevant low-energy scale is expected to vary roughly like the Kondo
temperature of the effective impurity model with $\Gamma
(\omega)$ as medium, i.e.\ $T_0 \sim \exp\left(-\pi U/ 8\Gamma(0) \right)$. 
Since this scale also sets the width of the quasiparticle peak, whose height 
is on the other hand fixed to $1\over \pi \Gamma(0)$ by Friedel's sum rule, 
the weight in the quasi-particle peak is proportional to
$\exp\left(-\pi U/ 8\Gamma(0) \right)/\Gamma(0)$, which falls
exponentially with decreasing $\Gamma(0)$.  Since the Drude peak is
formed from quasiparticle excitations, the fall in the $\Gamma(0)$
also corresponds to a loss in Drude weight.

        The weight lost in the Drude peak as $T$ decreases is gained by both 
the mid-IR peak and the low-frequency incoherent part since we find the total 
spectral weight to be roughly constant in $T$.  The weight gained in the 
incoherent part is reflected by an increase in $C_{inc}$ and by the reduction
of the exponent $\alpha$.  We observe that for $T\alt T_0$ the exponent starts 
to deviate from its Fermi-liquid value $2$, indicating that additional weight
is building up in the tail of the low frequency peak. The temperature 
dependence of $D^*$ is basically the same as that of $D_0+C_{inc}$ 
(cf.\ Fig.~\ref{Drude_Gamma}) suggesting that there is a low-frequency 
incoherent contribution to the spectral spectral function, in addition to
the quasiparticle peak, due to non-Fermi-liquid nature of the problem in this 
parameter regime.  The remaining transferred weight is gained by the 
peak in the mid-IR region which is due to excitations between the occupied and 
unoccupied bands states.  Since the optical conductivity is constructed from a 
convolution of lattice spectral functions, the origin of this peak can be 
confirmed by inspecting the dispersion of these bands which is plotted in the 
inset to Fig.~\ref{Optical_Band}.  Here, the solid line is determined from 
the real-part of the poles of the lattice propagators, thus it represents 
the quasiparticle dispersion.  The location of the peaks in the single-particle 
spectra are denoted by the symbols.  The singly occupied $f$-level appears as 
an almost dispersionless band below the Fermi energy. 
The energy of the mid-IR peak coincides with that of the interband transition 
between the two quasiparticle bands below and above the Fermi energy at 
$\epsilon_k \approx 0$ (this value of $\epsilon_k$ is special since it
represents most of the points in the Brillouin zone of the infinite 
dimensional lattice.) 

        Both the Drude peak and the mid-IR peak in $\sigma(\omega)$ persist 
up to unusually high temperatures $T\approx 10 T_0$.  In the conventional 
Kondo picture, based upon the single-impurity model, the quasiparticle peak 
in the DOS, and hence both of these features in $\sigma(\omega)$, are expected 
to disappear at much lower temperatures relative to the Kondo scale.   The 
slow evolution of these features may be due to a crossover between two 
energy scales\cite{niki2,niki_prl}; $T_K$, the screening scale of the 
impurity model with the same parameters as the PAM, which is the temperature 
where screening starts, and $T_0$ the Wilson-Kondo scale of the lattice.
This screening regime is extended, over that in the SIAM, because of the 
reduction in $\Gamma(0)$.  I.e.\ as $T$ is lowered below $T_K$, the Kondo 
scale of effective impurity problem in the DMFA is self-consistently 
reduced.  Thus, the slow evolution of $D_0$, $C_{MIR}$ and the quasiparticle 
peak itself are direct consequences of exhaustion.

\subsection{Resistivity}

\begin{figure}[htb] 
\epsfxsize=3.0in
\epsffile{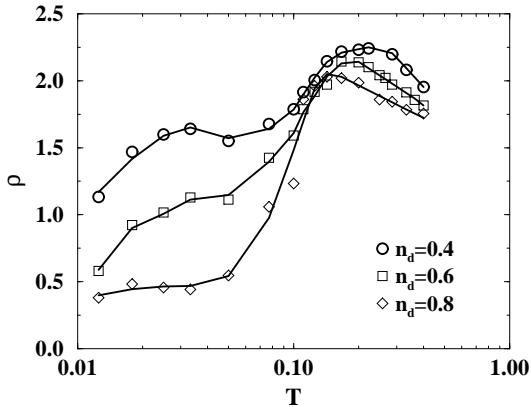}
\caption[Resistivity of the PAM]
{\em {The resistivity  vs. temperature calculated with QMC-MEM
when $U=1.5$, $V=0.6$, $n_f\approx 1$
for $n_d=0.4,\,0.6$ and $0.8$ (with $T_0=0.017,\,0.055$ and $0.149$,
respectively).  The solid lines are from fits to a 
polynomial.  At high temperatures, the resistivity has an impurity-like
log-linear regime which terminates maximum at $T=T_e$.  $T_e$ is non-universal
in that it is independent of $T_K$ or $T_0$ which increase with $n_d$;
whereas $T_e$ decreases with $n_d$.  When $n_d=0.4$ and $0.6$, $T_e$ 
corresponds to the temperature at which exhaustion becomes apparent 
as a dip in $\Gamma(\omega)$.  For $T<T_e$, the resistivity shows a 
non-universal crossover region (see text).  When $n_d=0.4$ and $0.6$,
at still lower temperatures 
the resistivity begins to drop quickly at $T\approx T_0$, indicating the 
onset of Fermi-liquid formation and coherence.  However for $n_d=0.8$ there 
is no such drop consistent with the proximity to the insulating regime 
($n_f=n_d=1$).
}}
\label{resistivity}
\end{figure}

The resistivity $\rho=1/\lim_{\omega \to 0} \sigma(\omega)$ also displays
anomalies due to exhaustion. $\rho$ versus $T$ from a QMC-MEM calculation
is shown in Fig.~\ref{resistivity} in a semilog plot for three different 
conduction band fillings $n_d$.  

We can see the interplay between the different energy scales of the PAM 
in Fig.~\ref{resistivity} and distinguish three distinct characteristic 
temperature regions for the resistivity.  
i) 
A region of log-linear 
resistivity which occurs at high temperatures $T\agt T_K$ where the 
single impurity physics is dominant and the correlated sites act like 
independent (incoherent) scattering centers. As a result, the resistivity 
displays the log-linear behavior of a dilute magnetic alloy. 
ii) 
As the temperature is lowered, the resistivity reaches a maximum value
at $T=T_e$.  Note that $T_e$ is non-universal in that it is independent 
of $T_K$ or $T_0$ which increase with $n_d$; whereas $T_e$ decreases with 
$n_d$.  At least for the $n_d=0.4$ and $0.6$ datasets, $T_e$ corresponds 
to the temperature at which exhaustion first becomes apparent as a dip 
in $\Gamma(\omega)$.  
As the temperature is lowered further, $T<T_e$, a pronounced dip begins
to develop in $\Gamma(\omega)$, and is accompanied by a rapid
drop in the resistivity.  Then, as the decline in $\Gamma(0)$ slows, the 
resistivity falls more slowly.  Since the impurity scattering rate is 
expected to vary like $\sim \ln\left(\frac{T}{\sqrt{\Gamma(0)U}}
\exp\left(-8U/\pi\Gamma(0)\right)\right)$\cite{hewson}, the  resistivity 
can fall with 
decreasing $\Gamma(0)$ {\em{even though the system has not begun to form a 
Fermi liquid}}.  Partial Fermi-liquid formation can also contribute to the
drop; however, this effect would be suppressed for small $n_d$ by 
exhaustion and the concurrent reduction of the effective Kondo scale.  
For $n_d\approx 0.6$ it is uncertain which of the two mechanisms has 
the dominant contribution 
to the drop in the resistivity at the top of the crossover regime. Also 
it is unclear why the crossover regime becomes less pronounced as $n_d$ 
decreases.  However, it seems likely that the drop for small $n_d$, where 
exhaustion is most pronounced, is due the  drop in the the scattering rate.  
Whereas for  $n_d\to 1$, where there 
is no exhaustion (the Kondo scale $T_0$ is actually 
enhanced relative to the impurity scale $T_K$\cite{rice_ueda,symmpam}) 
it seems likely
that the initial drop is due to partial Fermi-liquid formation.
iii)
Finally, as the temperature is reduced further, for conduction band
fillings $n_d=0.4$ and $0.6$, the resistivity begins to drop quickly at
$T\approx T_0$, indicating the onset of Fermi-liquid formation at $T_0$ for 
these fillings.  However for $n_d=0.8$ there is no indication of this for 
temperatures well below $T_0$.  When $n_f=n_d=1$, the PAM forms an
insulating gap of width $\approx T_0$\cite{symmpam} (which is resonantly 
enhanced over the impurity scale $T_K$).  Thus, 
for $n_d\alt 1$, we expect a drop in the resistivity to occur when
the temperature equals roughly the energy difference between the chemical 
potential and the bottom of the gap.

\section{Comparison with Experiment}

        Exhaustion should be most prevalent in heavy fermion materials
with low carrier concentration or a low density of conduction band
states at the Fermi surface.  One such material, for which an extensive
study of the optical conductivity has been done, is Yb$_4$As$_3$ with
a Kondo scale of $T_0\approx 40$K \cite{Yb4As3}.  This material has 
strongly temperature-dependent Drude and MIR (at roughly $0.4$ eV) peaks for 
temperatures ranging from 39K to 320K, as predicted by our model calculation.  
In addition, there is a very significant weight transfer from the Drude peak 
to the MIR peak as the temperature increases over this range\cite{alternate}.  
There is also some evidence for a Drude weight which drops with lowering 
the temperature below $T_e$ in a typical paramagnetic HF 
material like CeAl$_3$\cite{degorgiCeAl3}.  Here, the transfer of weight 
is far less apparent, but may be inferred from the analysis of the data,
where the inverse of the Drude weight (measured by fitting the optical 
conductivity to a Drude form at low frequencies) shows an increase with 
lowering $T$ at $\omega =0$.  
However, in this material, the Drude weight vanishes quickly as the 
temperature is increased to $T=10K\approx 3T_0$.  Thus, exhaustion, and the 
concomitant protracted evolution of $D(T)$ would seem to be less evident 
in this material than it is in the low carrier system Yb$_4$As$_3$.  
Unfortunately, there is not enough 
optical data available for paramagnetic HF systems to determine whether 
the drop in the Drude weight and other feature associated with exhaustion 
are generic features or even to determine their dependence on the f-level 
degeneracy and the conduction band filling.  We believe that a systematic  
study of the optical properties of low carrier density HF systems could 
shed significant light on the importance of exhaustion in these materials.

\section{Conclusion}

        We have performed a QMC-MEM simulation of the infinite-dimensional
Periodic Anderson Model to study the optical conductivity in the metallic 
regime.  We see several anomalies, including: (i) a Drude peak which persist 
up to anomalously high temperatures $T\alt 10 T_0$,  (ii) a MIR peak due 
to interband transitions across the Fermi energy, which also 
persists up to high temperatures $T\alt 10 T_0$,  (iii) a significant loss 
of Drude weight and f-d hybridization as $T \to 0$.  We interpreted these 
results in terms of Nozi$\grave{e}$res' exhaustion picture. Finally, we note 
that since optical measurements are far less surface sensitive than 
photoemission measurements,  the presence of these features in experimental 
spectra could serve as less controversial evidence for exhaustion in 
heavy Fermion systems.

\acknowledgments  
We would like to acknowledge useful conversations with 
J.W.\ Allen,
A.\ Arko,
L.\ Degiorgi,
M.\ Hettler,
J.\ Joyce,
H.R.\ Krishnamurthy,
and G.\ Thomas.
This work was supported by the National Science Foundation grants 
DMR-9704021, DMR-9357199, and the Ohio Supercomputing Center.
JKF was supported by the Office of Naval Research YIP N000149610828.

\end{document}